\documentclass[%
 reprint,
superscriptaddress,
 amsmath,amssymb,
 aps,
 pre
]{revtex4-2}
\usepackage[pdftex]{graphicx}
\usepackage{dcolumn}
\usepackage{bm}

\usepackage{amsfonts, amsthm, amssymb}
\usepackage{color}
\usepackage{ascmac}
\usepackage{url}
\usepackage{ulem}
\usepackage{bm}
\usepackage{bbold}
\usepackage{amsmath}
\usepackage{algorithm}
\usepackage{algpseudocode}
\usepackage{braket}
\usepackage{multirow}
\usepackage{subfiles}
\usepackage{cprotect}
\usepackage{epsfig}
\usepackage{xspace}
\usepackage{tipa}
\usepackage{overpic}

\begin{document}

\title{Tracking Cluster Continuity and Dynamics in Time-Series Data: Application to Chromatin Polymer Simulations}
\author{Ryo Nakanishi}
\affiliation{
 Graduate School of Arts and Sciences,
 The University of Tokyo,
 Komaba, Meguro-ku, Tokyo 153-8902, Japan
}

\author{Koji Hukushima}
\affiliation{
 Graduate School of Arts and Sciences,
 The University of Tokyo,
 Komaba, Meguro-ku, Tokyo 153-8902, Japan
}
\affiliation{
Komaba Institute for Science, The
University of Tokyo, 3-8-1 Komaba, Meguro-ku, Tokyo 153-8902, Japan
}
\date{\today}

\begin{abstract}
This study presents an enhanced method for analyzing cluster dynamics, with a particular focus on tracking clusters' continuity over time using time-series data from molecular dynamics (MD) simulation. The proposed method was applied to spatio-temporal cluster data obtained from a non-equilibrium MD simulation of a chromatin polymer model. 
In this model, clusters are formed on the polymer by binding molecules that stochastically and temporarily bind to the polymer segments at finite rates.
Our analysis successfully tracked the dynamics of clusters, including merging and splitting events, and revealed that clusters exhibit a percolation transition in both spatial and temporal domains. This suggests that clusters in the chromatin polymer model can persist even under finite rates of attractive interactions, demonstrating that the method can capture complex cluster dynamics over time. 
\end{abstract}

\maketitle

\section{Introduction}
Molecular dynamics (MD) simulations play an increasingly essential role in various research fields, such as materials science and biological science.
The rapid growth of computing power has made it possible to obtain large amounts of simulation data, making it more important than ever to fully exploit the information from these simulation data.
For example, in MD simulations of biomolecular systems, which are the subject of this study, it is possible to measure not only global statistics, such as the radius of gyration, but also the time-dependent evolution of the microscopic state of the system, including the positions and momenta of individual polymer segments and even solvent molecules, which are often difficult to measure experimentally. However, the microscopic states obtained in simulations contain a large number of degrees of freedom, and it is a non-trivial problem to determine which information is most meaningful to extract from them. 

In this context, clustering algorithms are widely used to analyze the resulting MD simulation data, such as classifying sampled conformations over trajectories\cite{shaoClusteringMolecularDynamics2007,pengClusteringAlgorithmsAnalyze2018} or identifying cluster structures at specific timesteps\cite{Brackley2017,Schneider2020}. 
For this purpose, clusters in snapshots of MD simulations at each timestep are usually determined by thresholding the distance between neighboring atoms or molecules, and algorithms implemented in simulation software such as LAMMPS\cite{LAMMPS2022} or VMD\cite{VMD_HUMP96} are widely used. 
However, beyond identifying clusters at a given snapshot in an MD simulation, it is also important to quantitatively evaluate the dynamics and evolution of these cluster structures over time. 
To address this, we developed a general method to track how cluster structures evolve in time by identifying key events such as cluster creation, annihilation, merging, or splitting from MD simulation trajectories. 
This method can be applied not only to MD simulation data but also to any other time-series data that exhibit cluster structures at each timestep.

To demonstrate the utility of this method,  we applied it to simulations of a biological polymer system, specifically a coarse-grained polymer chromatin model, to evaluate the time evolution of cluster structures on the chromatin. 
Through this analysis, we evaluated cluster stability in both the temporal and spatial (polymer) directions, identifying percolation transitions in both cases. This reveals that certain clusters can survive indefinitely despite finite rates of biding and unbinding in the model. In particular, we found that stable clusters emerge below a critical finite rate that, once formed, can persist for arbitrarily long times, indicating a transition from transient to persisting structures. These observations demonstrate the ability of our method to capture complex dynamical behaviors and to provide direct insight into critical percolation phenomena in the system. 

The structure of this paper is as follows: In Sec.~\ref{sec:methods},  we describe the polymer model used as an example in the MD simulations and introduce the proposed clustering method, which is the main contribution of this study. Sec.~\ref{sec:results} presents the results of the clustering analysis, focusing on percolation transitions in both the temporal direction and along the polymer. Finally, Sec.~\ref{sec:summary} provides a discussion and summary of this study. 

\section{Simulation model and proposed method}
\label{sec:methods}
In this section, we first introduce the simulation model used in our study. Next, we describe the clustering method used to analyze individual simulation timestep. We then explain the procedure for linking cluster structures across timesteps to perform a time-series analysis, allowing us to track the evolution of cluster structures and their cluster lineages over time.

\subsection{Coarse-grained polymer model of chromatin}
\label{sec:model}
We perform MD simulations and analyze the resulting time-series data using the clustering method described in the following subsection. The simulated system is based on a coarse-grained model of chromatin, the strings and binders switch (SBS) model\cite{Barbieri2012,Barbieri2013-jg}. 
The SBS model consists of a self-avoiding walk (SAW) polymer and binding molecules that can multivalently bind to monomers, which are polymer segments.
The polymer represents chromatin, a complex of DNA and histone proteins\cite{Cortini2016,Maeshima2021}.
This model has been proposed to explain the mechanism and behavior of cluster structures formed on chromatin.
In recent studies of the SBS model, the binding molecules can switch between bindable and unbindable states at a specific switching rate\cite{Brackley2017, Brackley2021, Semeraro2023}. 
This state-switching mechanism is crucial for preventing the entire polymer from collapsing into a compact globule state, instead allowing for the formation of finite-size clusters. These clusters represent a localized aggregation of monomers, which is an important characteristic of chromatin behavior in the model.  

The SAW polymer is modeled using the Kremer-Grest approach, where adjacent segments interact through the finitely extensible nonlinear elastic (FENE) potential\cite{kremerDynamicsEntangledLinear1990}, defined by
\begin{equation}
U^{\mathrm{FENE}}(r)= \begin{cases}
-\frac{1}{2} k R_{0}^{2} \ln \left(1-\left(\frac{r}{R_{0}}\right)^{2}\right), & r \leq R_{0},  \\
\infty, & r>R_{0},
\end{cases}
\end{equation}
where $R_0$ is the typical length scale and $k$ is a coupling constant.
The binding molecules can switch between two states, active and inactive, at a constant rate $\alpha$. The state switching corresponds to the chemical modifications of the binding molecules\cite{Brackley2017}. 
The binding molecules in the active state interact with polymer segments through an attractive Lennard-Jones (LJ) potential with a cutoff $r_c$, given by
\begin{equation}
\begin{aligned}
&U_{\mathrm{attractive}}(r)
\\
=& \begin{cases}
4 \varepsilon \Big[\left(\frac{\sigma}{r}\right)^{12}
   -\left(\frac{\sigma}{r}\right)^{6}
   -\left(\frac{\sigma}{r_{c}}\right)^{12}
   + & \left(\frac{\sigma}{r_{c}} \right)^{6} \Big], 
   \\
   & r < r_c = 1.8 \sigma, \\
0, & r \geq r_c = 1.8 \sigma,
\end{cases}
\end{aligned}
\end{equation}
where $\varepsilon$ defines the depth of the potential well, and $\sigma$ is the distance at which the potential becomes zero. In our simulations, the value of $\varepsilon$ is set such that the depth of the potential is $4 k_\mathrm{B}T$, following the previous study\cite{Brackley2017}.
The other particles, including polymer segments and inactive binding molecules, interact through a purely repulsive Weeks-Chandler-Andersen (WCA) potential \cite{WCA}, given by
\begin{equation}
U_{\mathrm{WCA}}(r)= \begin{cases}4 \varepsilon\left[\left(\frac{\sigma}{r}\right)^{12}-\left(\frac{\sigma}{r}\right)^{6}\right]+\varepsilon, & r<2^{1 / 6} \sigma,\\ 0, & r \geq 2^{1 / 6}  \sigma. \end{cases}
\end{equation}

Each particle follows Langevin dynamics, expressed by the equation: 
\begin{equation}
m \dot{\boldsymbol{v}}(t)=-\nabla V(\boldsymbol{r})-\gamma \boldsymbol{v}(t)+\boldsymbol{\xi}(t),
\end{equation}
where $m$ is the particle mass, $\gamma$ is the damping coefficient, and $\boldsymbol{r}$ and $\boldsymbol{v}$ are the position and velocity vectors, respectively. The random noise element $\xi_i(t)$ in the noise vector $\boldsymbol{\xi}(t)$ at temperature $T$ follows the fluctuation-dissipation relation: 
\begin{equation}
\langle\xi_i(t) \xi_j\left(t^{\prime}\right)\rangle=2 \mathrm{k}_\mathrm{B} T \gamma \delta_{ij}\delta\left(t-t^{\prime}\right),
\end{equation}
where $k_\mathrm{B}$ is the Boltzmann constant. 
Our simulation code is based on LAMMPS\cite{LAMMPS2022}, with an additional external code implemented to handle the state switching of the binding molecules. 
For a given rate $\alpha$, the state of each binding molecule is updated every $M$ timesteps with a probability $p$, which is given by 
\begin{equation}
    \begin{aligned}
    p =1-e^{-\alpha \Delta T},
    \end{aligned}
\end{equation}
where $\Delta T = M d t$ with the timestep width $dt$. This probability represents the stochastic switching process of the binding molecules during the time interval $\Delta T$.

\subsection{Method for identifying clusters and their dynamics}
As reported in a previous study\cite{Brackley2017}, polymer segments and active binding molecules form finite-size clusters, with the average size at steady state determined by the switching rate. In contrast,  we focus on tracking the detailed evolution of these finite-size clusters over time by identifying cluster lineages from the time-series data obtained from the MD simulations described in the previous subsection. In this subsection, we present the method for achieving this detailed analysis.  

The numerical approach to evaluate the time evolution of the clusters consists of the following two steps. 
First, clusters are identified from snapshots taken every $n$ timesteps during the MD simulations. A standard clustering method is used, in which polymer segments in close proximity to a common active binding molecule are classified as belonging to the same cluster. 
A unique cluster label is assigned to each identified cluster, and all polymer segments and binding molecules belonging to that cluster are marked with the same cluster label. 
The threshold distance is set as $d_\mathrm{c} = 3\sigma$. If multiple polymer segments are within a distance $d_\mathrm{c}$ of a specific active binding molecule, all of these polymer segments are classified as belonging to the same cluster. 
In addition, cluster labels are unified if two clusters share polymer segments. In other words, if the $i$th and $j$th polymer segments are close to one active binding molecule, and the $j$th and $k$th segments are close to a different active binding molecule, then all three segments are considered to be part of the same cluster. 

Second, we analyze the time evolution of the cluster structures by tracking how the cluster components, specifically the active binding molecules in this case, are inherited between clusters in consecutive simulation snapshots. 
After forming the clusters, they undergo coalescences, fissions, and annihilation events. 
Since attractive interactions exist only between active binding molecules and polymer segments, we define these events of cluster dynamics by comparing clusters at consecutive timesteps and classifying how the active binding molecules are inherited between clusters.

The ``creation" of a cluster is defined as the formation of a cluster where none of its active binding molecules were in the vicinity of any polymer segments at the previous timestep.
The ``annihilation" of a cluster is defined as the inverse process of the creation, where all active binding molecules in a cluster at a specific timestep leave the vicinity of the polymer in the subsequent timestep. 
The ``coalescence" of clusters occurs when active binding molecules from different clusters at the previous timestep belong to the same cluster at the subsequent timestep.
Conversely, the ``fission" of a cluster is defined as the reverse process of coalescence, where active binding molecules in a single cluster split into different clusters at the subsequent timestep.

\begin{figure}[h]
  \includegraphics[width=\linewidth]{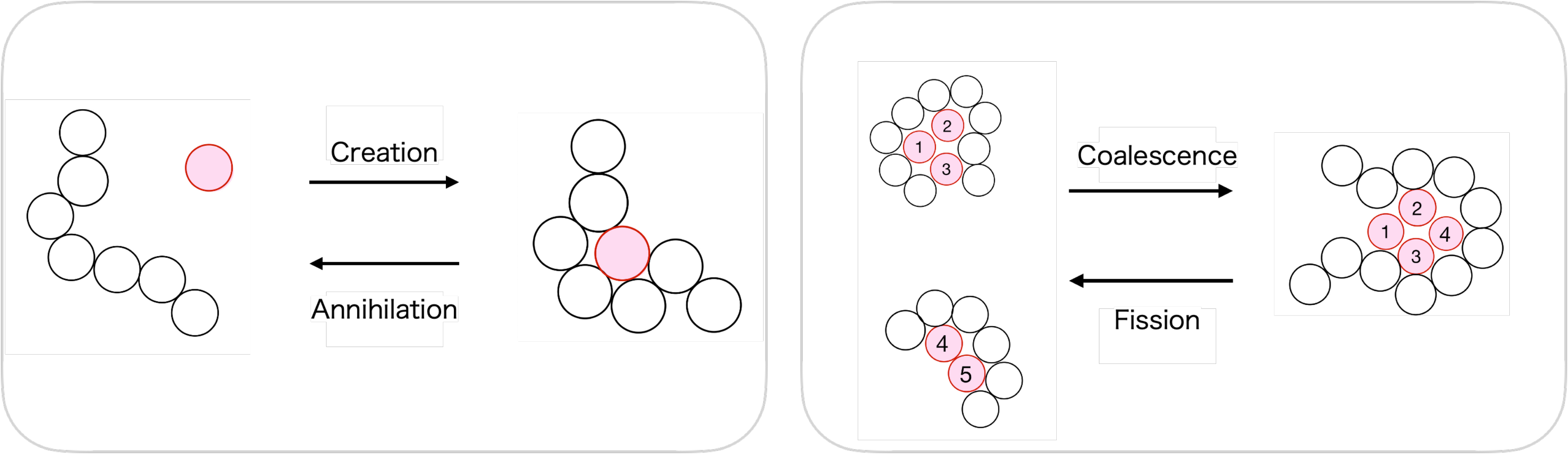}
  \caption{Four key cluster dynamics events. The white and red circles represent polymer segments and active binding molecules, respectively. The index numbers of the active binding molecules in the right figure indicate the movement of each molecule. 
  Left: Creation refers to the formation of a new cluster when active binding molecules come into the vicinity of polymer segments. Annihilation is the dissolution of a cluster when all active binding molecules move away from the polymer segments. Right: Coalescence represents the merging of two or more clusters into one when active binding molecules from distinct clusters at the previous timestep come together. Fission is the splitting of a single cluster into two, as binding molecules disperse into separated clusters.}
  \label{fig:events}
\end{figure}

These cluster dynamics events are presented in Fig.\ref{fig:events}.
Considering these events, we also define a ``cluster lineage" as the set consisting of a created cluster and all its descendant clusters.
The ``duration time" of a cluster lineage is defined as the time interval between the creation of a cluster and the disappearance of all its descendant clusters.

\section{Numerical Results}
\label{sec:results}
In this study, we performed MD simulations using the polymer model described in Sec.~\ref{sec:model}. The polymer consists of $1000$ segments, and the number of binding molecules is $N_\mathrm{BM} = 400$. We adopted LJ dimensionless reduced unit, where $m$, $\sigma$, $\varepsilon$, $\mathrm{k}_\mathrm{B}$ and the time unit $\tau_0 = \sqrt{m\sigma^2/\varepsilon}$ were set to unity. The linear size of the simulation box is $L = 100$, and the Langevin dynamics was simulated using the velocity Verlet method with a timestep of $dt = 0.01$. 
To estimate the typical timescale of this physical system, we used simple dimensional analysis. From the diffusion constant $D = 1$ and the density of binding molecules $\rho = 400/100^3$, the reciprocal of the typical timescale $\tau$ is estimated as $1/\tau = \rho^{2/3} D = 5\times 10^{-3}$. Based on this estimation, it is reasonable to set the switching rate of the binding molecules around this level. In our MD simulations, the switching rate $\alpha$ was controlled within the range of $2 \times 10^{-3} < \alpha < 5 \times 10^{-3}$.  Starting from randomly initialized polymer conformations and binding molecule positions, a relaxation period of $5\times 10^4$ steps was typically employed, followed by a measurement phase of $T=10^7$ steps.  

\subsection{Percolation transition in the time direction}
\begin{figure}[h]
  \includegraphics[width=\linewidth]{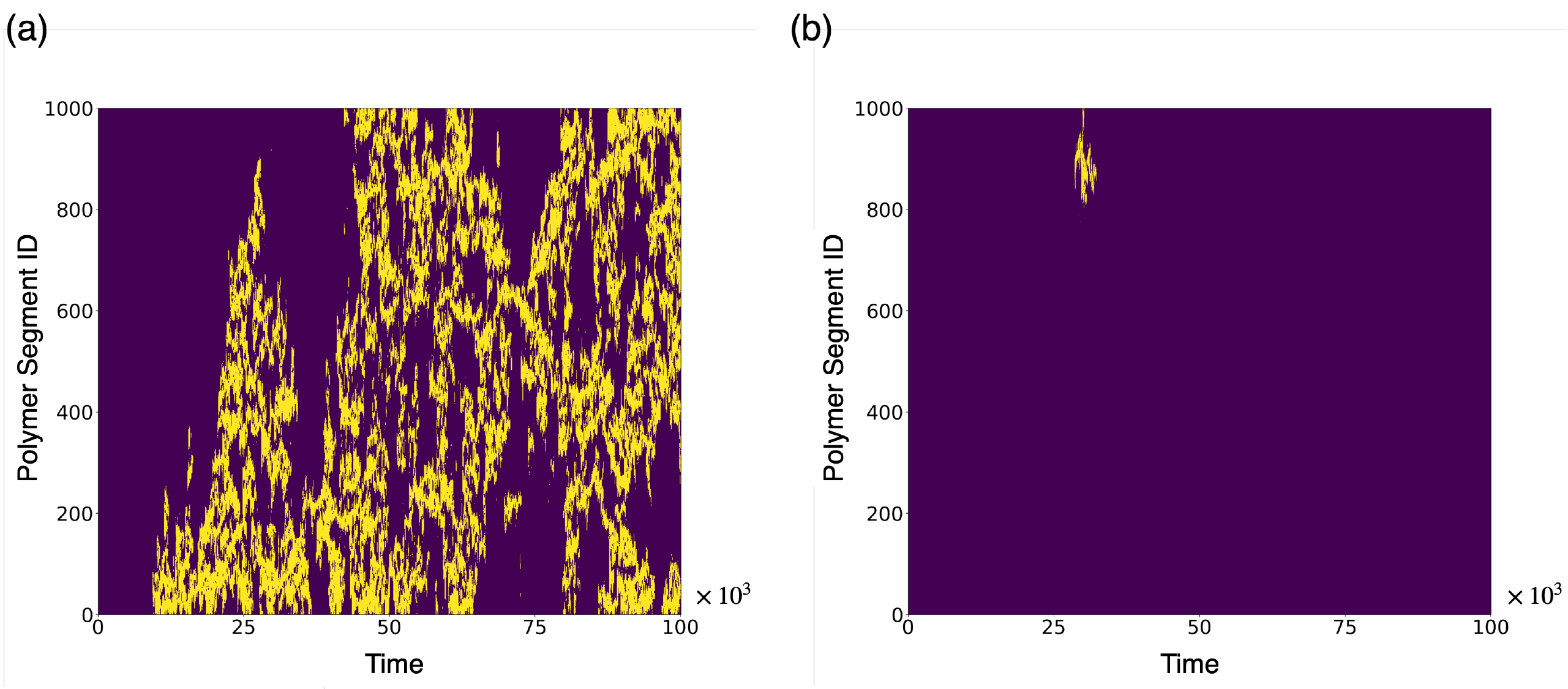}
  \caption{Examples of cluster lineages obtained by cluster analysis displayed as kymographs. Yellow cells indicate polymer segments that belong to a cluster lineage. The switching rate $\alpha$ is given for each example: (a) $\alpha = 2 \times 10^{-3}$, (b) $\alpha = 5 \times 10^{-3}$. }
  \label{fig:kymograph}
\end{figure}

From the time-series data obtained in the MD simulations, we extracted the cluster lineage for each value of $\alpha$. 
Figs.~\ref{fig:kymograph} show typical examples of cluster lineages for two different values of $\alpha$, presented as kymographs in the time and segment-ID space.  
The kymographs reveal that for small values of $\alpha$, cluster lineages appear to percolate in both the time and the polymer segment-ID directions. In contrast, for large values of $\alpha$, the cluster lineages are short-lived and do not percolate in either direction. 
This observation suggests the existence of an infinite-scale cluster lineage at a non-zero switching rate, corresponding to a percolation transition in an anisotropic system with time and segment-ID directions. Hence, this behavior is analogous to directed percolation\cite{Hinrichsen2000,HenkelHinrichsenLubeck2008} with respect to $\alpha$.

In the simulation, the spatio-temporal domain is characterized by two parameters: the maximum simulation time $T$ in the time direction and the polymer length $L$ in the segment direction. 
 To systematically study the possible percolation in the spatio-temporal domain, finite boxes are defined with a time window $\Delta t$ in the time direction and a segment width $\Delta l$ in the segment direction. We then examine whether percolation events occur within each box. By analyzing all such boxes across the entire $TL$ domain, the empirical probability of percolation events can be determined. First, we introduce the percolation rate of a cluster lineage, $r_\parallel(\alpha,\Delta t)$, for a given $\alpha$, as the probability of percolation in the time direction within boxes with a finite time window $\Delta t$ and the entire segment width $L$. This represents the probability that a cluster lineage persists over a time interval of $\Delta t$ within the box. 
 
\begin{figure*}[]
\begin{overpic}[width=0.32\linewidth]{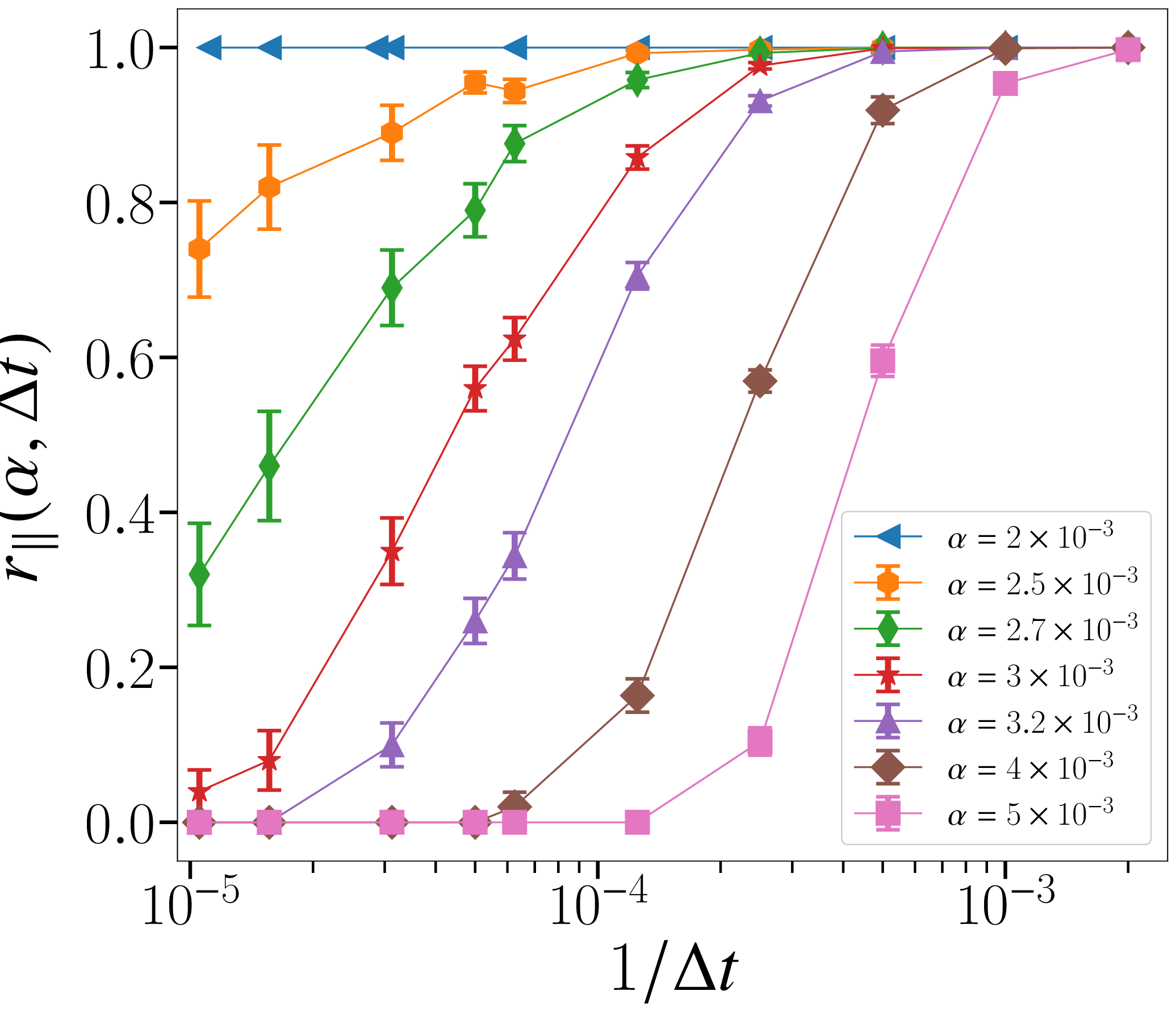}
   \put(18,75){(a)}
\end{overpic}
\begin{overpic}[width=0.32\linewidth]{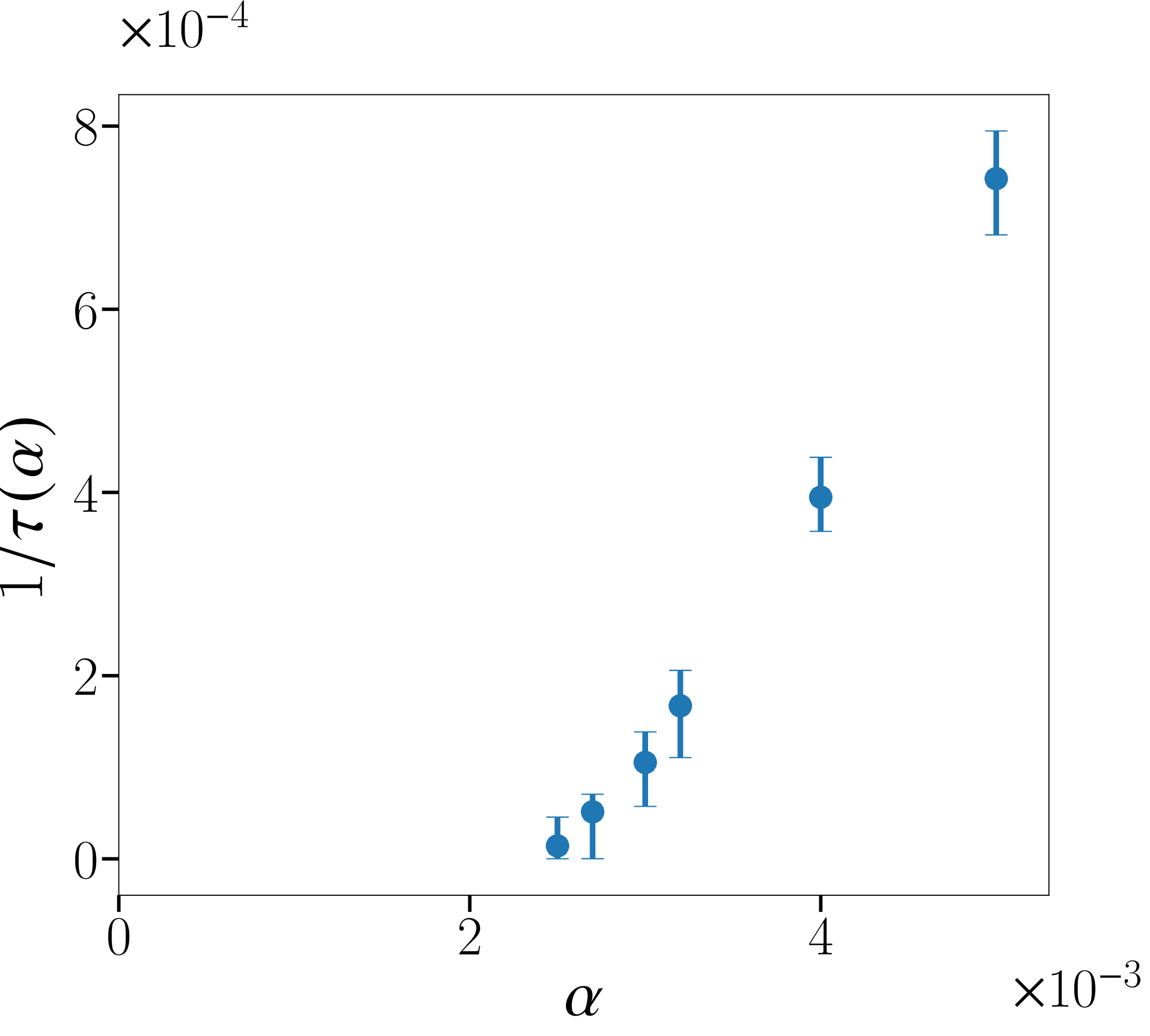}
   \put(15,70){(b)}
\end{overpic}
\begin{overpic}[width=0.33\linewidth]{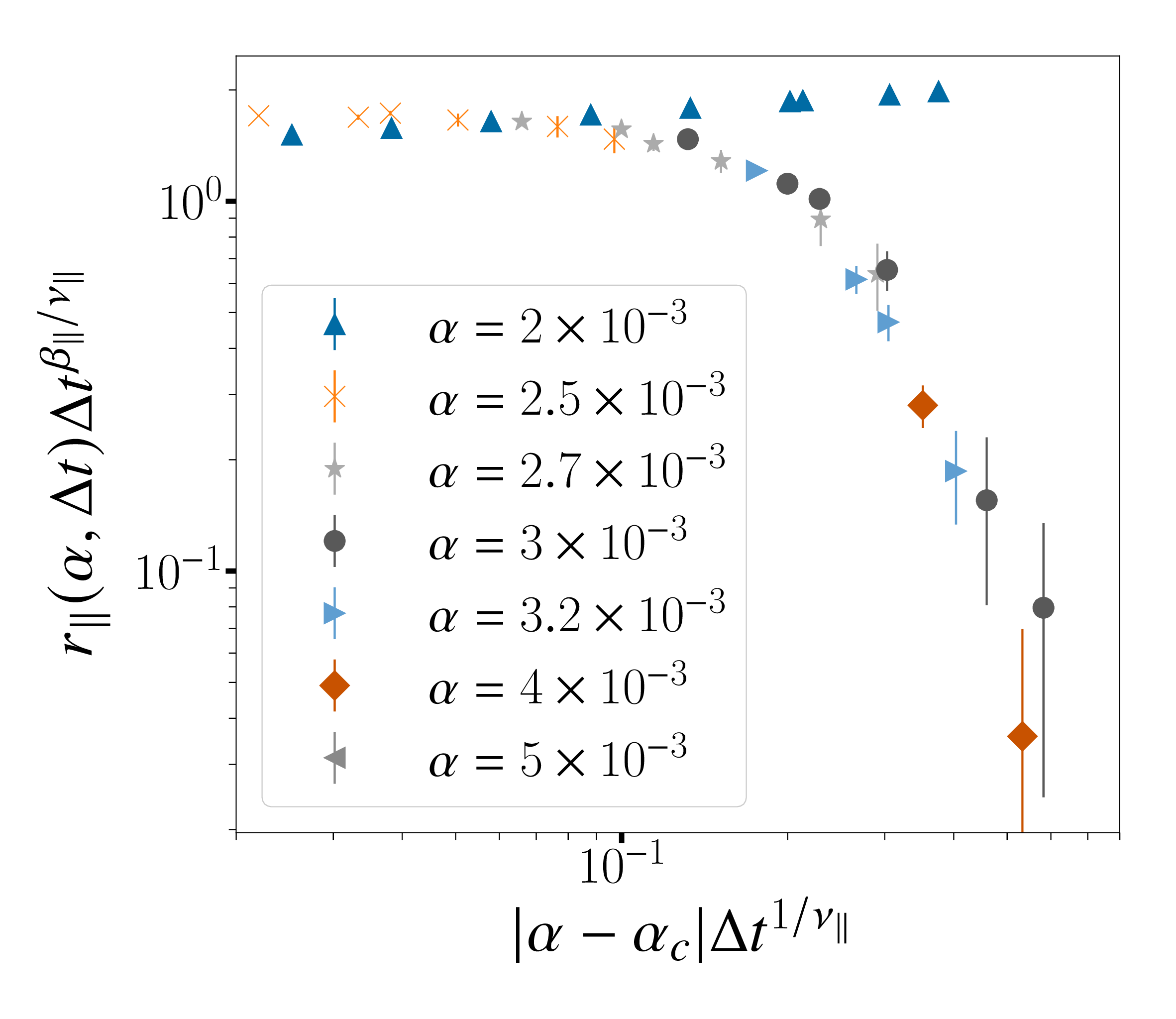}
   \put(22,69){(c)}
\end{overpic}
  \caption{(a) Percolation rate of cluster lineage for different values of the switching rate $\alpha$ as a function of the inverse window $\Delta t$ in the time direction. (b) Inverse characteristic timescale of cluster lineages as a function of the switching rate $\alpha$, defined as the time when the percolation rate is 0.8.  The error bars in (b) are estimated from linear interpolation of the error bars in (a). (c) Finite-time scaling plot of the percolation rate with $\alpha_c=2.4\times 10^{-3}$, $1/\nu_\parallel = 0.6$ and $\beta_\parallel=0.1$. The error bars in (a) and (c) represent the standard error, evaluated from $50$ independent simulations for $\alpha < 4 \times 10^3$ and $20$ independent simulations for $\alpha = 4 \times 10^3$ and $\alpha = 5 \times 10^3$. 
  }
\label{fig:perc-in-time}
\end{figure*}

Fig.~\ref{fig:perc-in-time}(a) shows the percolation rate for several values of $\alpha$ as a function of the inverse time window $\Delta t^{-1}$. When $\Delta t$ is small, the percolation rate is close to $1$ and decreases with increasing $\Delta t$, depending on the value of $\alpha$. Percolation is considered to occur when this rate remains non-zero, even for an infinitely large $\Delta t$. 
Therefore, in the case of a second-order transition, the $\alpha$ dependence of $r_\parallel(\alpha,\Delta t)$ in the infinite $\Delta t$ limit is expected to behave as follows: 
 \begin{equation}
r_\parallel(\alpha, \Delta t =\infty)\propto\left\{\begin{array}{rr}
\left|\alpha-\alpha_c\right|^{\beta_\parallel},  &\alpha<\alpha_c \\
0,\quad \quad  \quad& \alpha>\alpha_c
\end{array}\right., 
\label{eqn:critical}
\end{equation}
where $\alpha_c$ is the critical switching rate for the percolation transition, and $\beta_\parallel$ is a critical exponent that takes a positive value.  
In practice, we evaluate the characteristic timescale $\tau(\alpha)$ for the duration of cluster lineages by determining the value of $\Delta t$ where $r_\parallel(\alpha,\Delta t) = 0.8$ for each $\alpha$. 
Interestingly, $\tau(\alpha)$ shows a tendency to diverge as $\alpha$ decreases, as shown in Fig.~\ref{fig:perc-in-time}(b). By extrapolating the plot of $1/\tau(\alpha)$ as a function of $\alpha$, we estimate the critical switching rate to be $\alpha_c \simeq 2.4 \times 10^{-3}$, at which $\tau(\alpha)$ becomes infinite. This indicates a cooperative phenomenon, supporting a percolation transition at a finite switching rate.  

Following the rough estimation of the critical switching rate above, we now perform the finite ``time" scaling analysis, i.e., the finite size scaling analysis in the time direction of the kymograph. This analysis aims to study the critical properties and critical exponents associated with the percolation transition. The scaling hypothesis for the percolation rate $r_\parallel(\alpha, \Delta t)$  as a function of $\alpha$ and $\Delta t$ is given as follows:
\begin{equation}
r_\parallel(\alpha, \Delta t)=\Delta t^{-\beta_\parallel/\nu_\parallel} f_\parallel\left(\left|\alpha-\alpha_c\right| \Delta t^{1/\nu_\parallel}\right),
\label{eqn:FSS}
\end{equation}
where $\nu_\parallel$ is the correlation-length exponent, and $f$ is a universal scaling function. Consistency with the asymptotic form of $r_\parallel(\alpha,\Delta t)$ in Eq.~(\ref{eqn:critical}) indicates that the scaling function behaves as $f_{\parallel}(x) \propto x^{\beta_\parallel}$ for large $x$, and $\nu_\parallel$ is positive. 
This scaling analysis used the previously estimated value of $\alpha_c$.  
As shown in Fig.~\ref{fig:perc-in-time}(d), the scaling plot reveals two branches that correspond to the values of $\alpha$ below and above $\alpha_\mathrm{c}$, and the data for different $\alpha > \alpha_\mathrm{c}$ values collapse reasonably well, with the scaling exponent values of $1/\nu_\parallel = 0.6$ and $\beta_\parallel = 0.1$.

\subsection{Percolation transition along the polymer direction}
In this subsection, we analyze the percolation transition along the polymer direction. 
The percolation rate along the polymer, $r_\perp(\alpha,\Delta l)$, is defined as the probability that a cluster lineage originating from a cluster at the first timestep of the window reaches both ends of the polymer ID window with a width of $\Delta l$ during the maximum period $T$ in our simulations.  

Similar to the percolation in the time direction, we assume that the critical behavior of $r_\perp(\alpha,\Delta l=\infty)$ follows: 
\begin{equation}
r_\perp(\alpha, \Delta l =\infty)=\left\{\begin{array}{rr}
\left|\alpha-\alpha_c\right|^{\beta_\perp},  &\alpha<\alpha_c, \\
0,\quad \quad  \quad& \alpha>\alpha_c, 
\end{array}\right.
\end{equation}
where $\alpha_c$ is the critical switching rate for the percolation transition along the polymer, and $\beta_\perp$ is the critical exponent for this direction. It is important to note that, in principle, the critical switching rate $\alpha_c$ along the polymer direction may differ from that in the time direction. Additionally, $\beta_\perp$ is introduced as a separate critical exponent from $\beta_\parallel$, reflecting the distinct nature of the anisotropic percolation.  

\begin{figure}[ht]

\begin{overpic}[width=\linewidth]{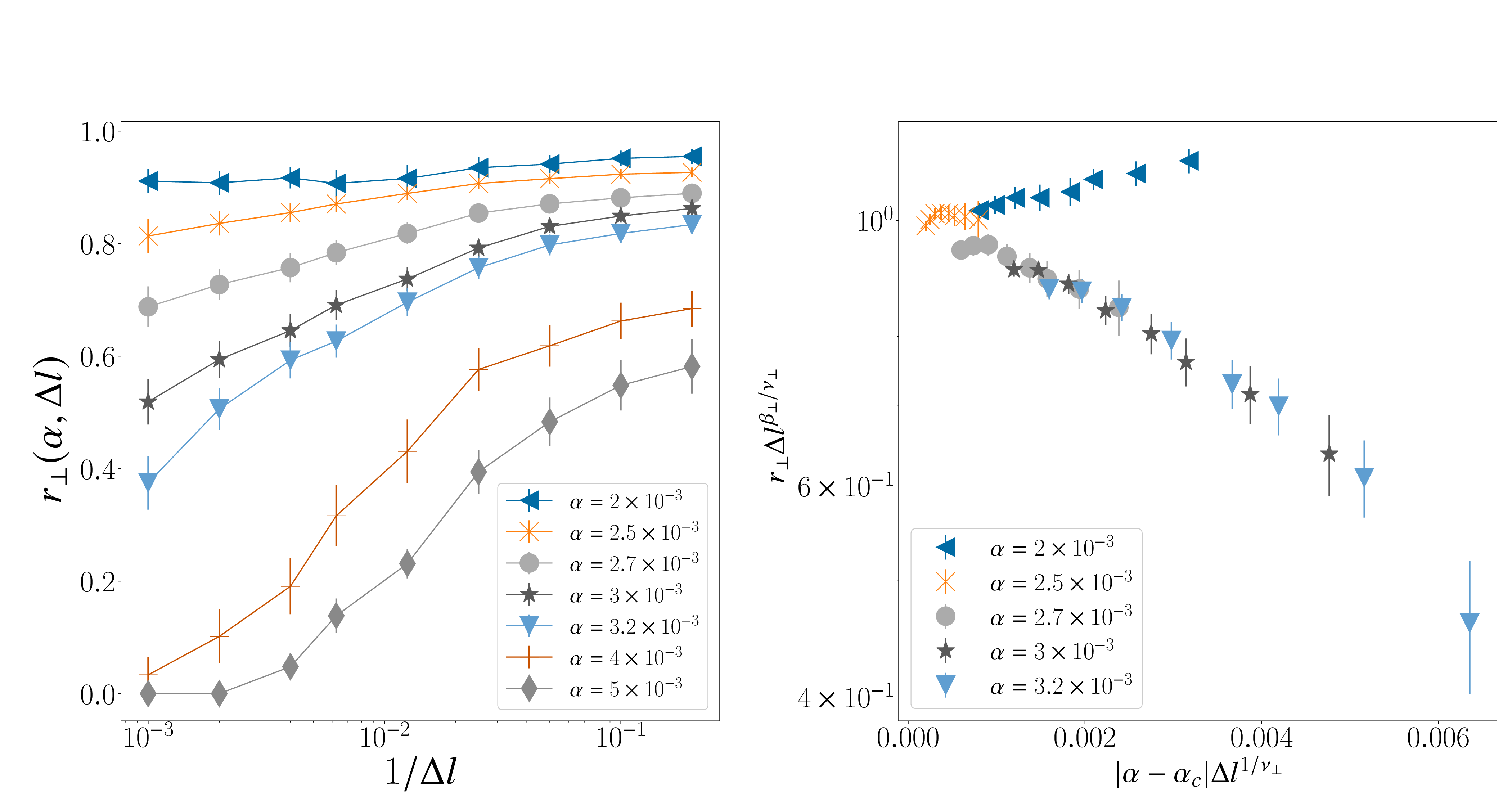}
    \put(0,45){(a)}
    \put(50,45){(b)}
\end{overpic}
  \caption{
  (a) 
  Percolation rate of cluster lineage in the polymer segment direction as a function of the inverse segment width $\Delta l$.  
  (b) Finite-size scaling plot of the percolation rate with parameters  $\alpha_c=2.4\times 10^{-3}$, $1/\nu_\perp = 0.3$ and $\beta_\perp = 0.1$.
  }
\label{fig:perc-in-length}
\end{figure}

We also conducted a finite-size scaling analysis for the percolation rate $r_\perp(\alpha,\Delta l)$. The scaling hypothesis for $r_\perp(\alpha,\Delta l)$ as a function of the switching rate $\alpha$ and the polymer length $\Delta l$ is given as follows: 
\begin{equation}
r_\perp(\alpha, \Delta t)=\Delta l^{-{\beta_\perp}/\nu_\perp} f_\perp\left(\left|\alpha-\alpha_c\right| \Delta l^{1/\nu_\perp}\right),
\end{equation}
where $\nu_\perp$ is the critical exponent for the correlation length along the polymer direction. 
The result, shown in Fig. \ref{fig:perc-in-length}(b), demonstrates two branches corresponding to values of $\alpha$ both below and above $\alpha_\mathrm{c}$ and a reasonable collapse of the data for $r_\perp(\alpha,\Delta l)$ onto a universal curve. This is obtained using the critical exponents $1/\nu_\perp=0.3$ and $\beta_\perp=0.1$, assuming the same critical switching rate $\alpha_c$ as for the percolation in the time direction. 

The results indicate that the percolation transition occurs at a common switching rate $\alpha_c$ in both the time and polymer directions. Furthermore, the percolation rate exponents, $\beta_\perp$ and $\beta_\parallel$, coincide. In contrast, the correlation length exponents, $\nu_\perp$ and $\nu_\parallel$, differ between the two directions, indicating the anisotropic percolation. As shown in Fig.~\ref{fig:kymograph}, the system can be considered a $1+1$ dimensional anisotropic system in time and space, while the polymer lies in three-dimensional space. Nevertheless, the evaluated critical exponents significantly differ from those of the universality class of directed percolation in $1+1$ dimensions. In particular, the dynamical exponent $z=\nu_\parallel/\nu_\perp$, which is typically close to $2$\cite{HenkelHinrichsenLubeck2008}, is found to be much smaller, approximately $0.5$, in our results. The cluster observed in this study may not arise from the directed percolation process, suggesting that the percolation transition belongs to a different universality class from directed percolations, which remains to be investigated further.  

\section{Summary and Discussion}
\label{sec:summary}
In this study, we performed MD simulations of the SBS model, a single polymer chromatin model. In this model, polymer segments attract each other indirectly through binding molecules, which switch between active and inactive states at a finite rate, controlling the strength of the attractive interactions. While it is clear that stable clusters of polymer segments emerge when no switching occurs, that is, when the switching rate is zero, this study addresses whether clusters with infinite lifetimes can form under finite switching rates.  

To investigate this, we developed a systematic method for extracting and tracking clusters on the polymer over time from the time-series data obtained in the MD simulations. This method identifies cluster dynamics, including creation, annihilation, coalescence, and fission events, and defines cluster lineage based on the inheritance of cluster components. 
Notably, this method to analyze cluster lineage and cluster events can be applied not only to the SBS model but also to other systems with time-series data that has cluster structure at each timestep.
We then evaluated the stability of clusters by analyzing the percolation rate of the obtained cluster lineage.
These lineages exhibit behavior similar to directed-percolation transitions, as seen in the kymographs. 

Our percolation analysis of the cluster lineages revealed that stable cluster lineage structures can emerge and persist over infinite timescales, even under non-equilibrium conditions where binding molecules repeatedly switch between active and inactive states, moving in and out of clusters. 
Experimentally, the formation of stable clusters on chromatin has been observed\cite{MAO2011295, Erdel2023}. From the perspective of the SBS model, these long-lived clusters observed in experiments suggest that the switching rate of binding molecules in biological systems is below the critical switching rate at which the percolation transition occurs. This comparison between the model and the experimental results supports the idea that the biological switching rates lie in a region where clusters can exist stably for long periods, i.e., within the percolation phase. 

In our simulations, cluster lineages also exhibit percolation along the polymer direction. 
However, this percolation behavior is inconsistent with experimental observations, where chromatin clusters are localized and fixed to specific sequences on the genome, playing functional roles. In biological systems, the absence of percolation along the polymer direction implies that the percolation observed in the simulations results from the simplified homopolymer model. 
In particular, the finite-size clusters observed in the simulation can move along the polymer, while experimentally observed clusters remain localized.  
The time-direction percolation, where clusters remain stable over long timescales, is consistent with experimentally observed stable chromatin clusters. However, the lack of percolation along the polymer in biological systems likely arises from heteromorphic properties such as sequence specificity or epigenetic marks that stabilize clusters on the chromatin polymer, preventing their movement along the polymer. 
Incorporating these features into the model could provide a more realistic representation of chromatin clustering and provide deep insights into the mechanism underlying their stability.

The clustering method proposed in this study plays a crucial role in systematically identifying and tracking cluster dynamics in time-series data. By enabling detailed analyses of cluster formation and lineage, this method provides a flexible framework for studying not only the chromatin-based SBS model but also other molecular systems with dynamic cluster structures. Its broad applicability makes it a valuable tool for advancing molecular dynamics analyses.

\section*{Acknowledgement}
This work was supported by MEXT as the Program for Promoting Research on the Supercomputer Fugaku (DPMSD, Project ID: JPMXP1020200307), JSPS KAKENHI Grant Number 23H01095, and JST Grant Number JPMJPF2221. One of the authors, RN, was supported by the WINGS-FMSP program at the University of Tokyo.
The authors thank the Supercomputer Center, the Institute for Solid State Physics, the University of Tokyo for the use of the facilities.

\bibliographystyle{apsrev4-2}
\bibliography{ref}

\begin{thebibliography}{18}%
\makeatletter
\providecommand \@ifxundefined [1]{%
 \@ifx{#1\undefined}
}%
\providecommand \@ifnum [1]{%
 \ifnum #1\expandafter \@firstoftwo
 \else \expandafter \@secondoftwo
 \fi
}%
\providecommand \@ifx [1]{%
 \ifx #1\expandafter \@firstoftwo
 \else \expandafter \@secondoftwo
 \fi
}%
\providecommand \natexlab [1]{#1}%
\providecommand \enquote  [1]{``#1''}%
\providecommand \bibnamefont  [1]{#1}%
\providecommand \bibfnamefont [1]{#1}%
\providecommand \citenamefont [1]{#1}%
\providecommand \href@noop [0]{\@secondoftwo}%
\providecommand \href [0]{\begingroup \@sanitize@url \@href}%
\providecommand \@href[1]{\@@startlink{#1}\@@href}%
\providecommand \@@href[1]{\endgroup#1\@@endlink}%
\providecommand \@sanitize@url [0]{\catcode `\\12\catcode `\$12\catcode
  `\&12\catcode `\#12\catcode `\^12\catcode `\_12\catcode `\%12\relax}%
\providecommand \@@startlink[1]{}%
\providecommand \@@endlink[0]{}%
\providecommand \url  [0]{\begingroup\@sanitize@url \@url }%
\providecommand \@url [1]{\endgroup\@href {#1}{\urlprefix }}%
\providecommand \urlprefix  [0]{URL }%
\providecommand \Eprint [0]{\href }%
\providecommand \doibase [0]{https://doi.org/}%
\providecommand \selectlanguage [0]{\@gobble}%
\providecommand \bibinfo  [0]{\@secondoftwo}%
\providecommand \bibfield  [0]{\@secondoftwo}%
\providecommand \translation [1]{[#1]}%
\providecommand \BibitemOpen [0]{}%
\providecommand \bibitemStop [0]{}%
\providecommand \bibitemNoStop [0]{.\EOS\space}%
\providecommand \EOS [0]{\spacefactor3000\relax}%
\providecommand \BibitemShut  [1]{\csname bibitem#1\endcsname}%
\let\auto@bib@innerbib\@empty
\bibitem [{\citenamefont {Shao}\ \emph {et~al.}(2007)\citenamefont {Shao},
  \citenamefont {Tanner}, \citenamefont {Thompson},\ and\ \citenamefont
  {Cheatham}}]{shaoClusteringMolecularDynamics2007}%
  \BibitemOpen
  \bibfield  {author} {\bibinfo {author} {\bibfnamefont {J.}~\bibnamefont
  {Shao}}, \bibinfo {author} {\bibfnamefont {S.~W.}\ \bibnamefont {Tanner}},
  \bibinfo {author} {\bibfnamefont {N.}~\bibnamefont {Thompson}},\ and\
  \bibinfo {author} {\bibfnamefont {T.~E.}\ \bibnamefont {Cheatham}},\ }\href
  {https://doi.org/10.1021/ct700119m} {\bibfield  {journal} {\bibinfo
  {journal} {Journal of Chemical Theory and Computation}\ }\textbf {\bibinfo
  {volume} {3}},\ \bibinfo {pages} {2312} (\bibinfo {year} {2007})}\BibitemShut
  {NoStop}%
\bibitem [{\citenamefont {Peng}\ \emph {et~al.}(2018)\citenamefont {Peng},
  \citenamefont {Wang}, \citenamefont {Yu}, \citenamefont {Gu},\ and\
  \citenamefont {Huang}}]{pengClusteringAlgorithmsAnalyze2018}%
  \BibitemOpen
  \bibfield  {author} {\bibinfo {author} {\bibfnamefont {J.-h.}\ \bibnamefont
  {Peng}}, \bibinfo {author} {\bibfnamefont {W.}~\bibnamefont {Wang}}, \bibinfo
  {author} {\bibfnamefont {Y.-q.}\ \bibnamefont {Yu}}, \bibinfo {author}
  {\bibfnamefont {H.-l.}\ \bibnamefont {Gu}},\ and\ \bibinfo {author}
  {\bibfnamefont {X.}~\bibnamefont {Huang}},\ }\href
  {https://doi.org/10.1063/1674-0068/31/cjcp1806147} {\bibfield  {journal}
  {\bibinfo  {journal} {Chinese Journal of Chemical Physics}\ }\textbf
  {\bibinfo {volume} {31}},\ \bibinfo {pages} {404} (\bibinfo {year}
  {2018})}\BibitemShut {NoStop}%
\bibitem [{\citenamefont {Brackley}\ \emph {et~al.}(2017)\citenamefont
  {Brackley}, \citenamefont {Liebchen}, \citenamefont {Michieletto},
  \citenamefont {Mouvet}, \citenamefont {Cook},\ and\ \citenamefont
  {Marenduzzo}}]{Brackley2017}%
  \BibitemOpen
  \bibfield  {author} {\bibinfo {author} {\bibfnamefont {C.~A.}\ \bibnamefont
  {Brackley}}, \bibinfo {author} {\bibfnamefont {B.}~\bibnamefont {Liebchen}},
  \bibinfo {author} {\bibfnamefont {D.}~\bibnamefont {Michieletto}}, \bibinfo
  {author} {\bibfnamefont {F.}~\bibnamefont {Mouvet}}, \bibinfo {author}
  {\bibfnamefont {P.~R.}\ \bibnamefont {Cook}},\ and\ \bibinfo {author}
  {\bibfnamefont {D.}~\bibnamefont {Marenduzzo}},\ }\href@noop {} {\bibfield
  {journal} {\bibinfo  {journal} {Biophys. J.}\ }\textbf {\bibinfo {volume}
  {112}},\ \bibinfo {pages} {1085} (\bibinfo {year} {2017})}\BibitemShut
  {NoStop}%
\bibitem [{\citenamefont {Schneider}\ \emph {et~al.}(2020)\citenamefont
  {Schneider}, \citenamefont {Meinel}, \citenamefont {Dittmar},\ and\
  \citenamefont {M\"{u}ller-Plathe}}]{Schneider2020}%
  \BibitemOpen
  \bibfield  {author} {\bibinfo {author} {\bibfnamefont {J.}~\bibnamefont
  {Schneider}}, \bibinfo {author} {\bibfnamefont {M.~K.}\ \bibnamefont
  {Meinel}}, \bibinfo {author} {\bibfnamefont {H.}~\bibnamefont {Dittmar}},\
  and\ \bibinfo {author} {\bibfnamefont {F.}~\bibnamefont
  {M\"{u}ller-Plathe}},\ }\href {https://doi.org/10.1021/acs.macromol.0c01315}
  {\bibfield  {journal} {\bibinfo  {journal} {Macromolecules}\ }\textbf
  {\bibinfo {volume} {53}},\ \bibinfo {pages} {8889–8900} (\bibinfo {year}
  {2020})}\BibitemShut {NoStop}%
\bibitem [{\citenamefont {Thompson}\ \emph {et~al.}(2022)\citenamefont
  {Thompson}, \citenamefont {Aktulga}, \citenamefont {Berger}, \citenamefont
  {Bolintineanu}, \citenamefont {Brown}, \citenamefont {Crozier}, \citenamefont
  {{in 't Veld}}, \citenamefont {Kohlmeyer}, \citenamefont {Moore},
  \citenamefont {Nguyen}, \citenamefont {Shan}, \citenamefont {Stevens},
  \citenamefont {Tranchida}, \citenamefont {Trott},\ and\ \citenamefont
  {Plimpton}}]{LAMMPS2022}%
  \BibitemOpen
  \bibfield  {author} {\bibinfo {author} {\bibfnamefont {A.~P.}\ \bibnamefont
  {Thompson}}, \bibinfo {author} {\bibfnamefont {H.~M.}\ \bibnamefont
  {Aktulga}}, \bibinfo {author} {\bibfnamefont {R.}~\bibnamefont {Berger}},
  \bibinfo {author} {\bibfnamefont {D.~S.}\ \bibnamefont {Bolintineanu}},
  \bibinfo {author} {\bibfnamefont {W.~M.}\ \bibnamefont {Brown}}, \bibinfo
  {author} {\bibfnamefont {P.~S.}\ \bibnamefont {Crozier}}, \bibinfo {author}
  {\bibfnamefont {P.~J.}\ \bibnamefont {{in 't Veld}}}, \bibinfo {author}
  {\bibfnamefont {A.}~\bibnamefont {Kohlmeyer}}, \bibinfo {author}
  {\bibfnamefont {S.~G.}\ \bibnamefont {Moore}}, \bibinfo {author}
  {\bibfnamefont {T.~D.}\ \bibnamefont {Nguyen}}, \bibinfo {author}
  {\bibfnamefont {R.}~\bibnamefont {Shan}}, \bibinfo {author} {\bibfnamefont
  {M.~J.}\ \bibnamefont {Stevens}}, \bibinfo {author} {\bibfnamefont
  {J.}~\bibnamefont {Tranchida}}, \bibinfo {author} {\bibfnamefont
  {C.}~\bibnamefont {Trott}},\ and\ \bibinfo {author} {\bibfnamefont {S.~J.}\
  \bibnamefont {Plimpton}},\ }\href@noop {} {\bibfield  {journal} {\bibinfo
  {journal} {Comput. Phys. Commun.}\ }\textbf {\bibinfo {volume} {271}},\
  \bibinfo {pages} {108171} (\bibinfo {year} {2022})}\BibitemShut {NoStop}%
\bibitem [{\citenamefont {Humphrey}\ \emph {et~al.}(1996)\citenamefont
  {Humphrey}, \citenamefont {Dalke},\ and\ \citenamefont
  {Schulten}}]{VMD_HUMP96}%
  \BibitemOpen
  \bibfield  {author} {\bibinfo {author} {\bibfnamefont {W.}~\bibnamefont
  {Humphrey}}, \bibinfo {author} {\bibfnamefont {A.}~\bibnamefont {Dalke}},\
  and\ \bibinfo {author} {\bibfnamefont {K.}~\bibnamefont {Schulten}},\
  }\href@noop {} {\bibfield  {journal} {\bibinfo  {journal} {Journal of
  Molecular Graphics}\ }\textbf {\bibinfo {volume} {14}},\ \bibinfo {pages}
  {33} (\bibinfo {year} {1996})}\BibitemShut {NoStop}%
\bibitem [{\citenamefont {Barbieri}\ \emph {et~al.}(2012)\citenamefont
  {Barbieri}, \citenamefont {Chotalia}, \citenamefont {Fraser}, \citenamefont
  {Lavitas}, \citenamefont {Dostie}, \citenamefont {Pombo},\ and\ \citenamefont
  {Nicodemi}}]{Barbieri2012}%
  \BibitemOpen
  \bibfield  {author} {\bibinfo {author} {\bibfnamefont {M.}~\bibnamefont
  {Barbieri}}, \bibinfo {author} {\bibfnamefont {M.}~\bibnamefont {Chotalia}},
  \bibinfo {author} {\bibfnamefont {J.}~\bibnamefont {Fraser}}, \bibinfo
  {author} {\bibfnamefont {L.-M.}\ \bibnamefont {Lavitas}}, \bibinfo {author}
  {\bibfnamefont {J.}~\bibnamefont {Dostie}}, \bibinfo {author} {\bibfnamefont
  {A.}~\bibnamefont {Pombo}},\ and\ \bibinfo {author} {\bibfnamefont
  {M.}~\bibnamefont {Nicodemi}},\ }\href@noop {} {\bibfield  {journal}
  {\bibinfo  {journal} {Proc. Natl. Acad. Sci. U.S.A.}\ }\textbf {\bibinfo
  {volume} {109}},\ \bibinfo {pages} {16173} (\bibinfo {year}
  {2012})}\BibitemShut {NoStop}%
\bibitem [{\citenamefont {Barbieri}\ \emph {et~al.}(2013)\citenamefont
  {Barbieri}, \citenamefont {Chotalia}, \citenamefont {Fraser}, \citenamefont
  {Lavitas}, \citenamefont {Dostie}, \citenamefont {Pombo},\ and\ \citenamefont
  {Nicodemi}}]{Barbieri2013-jg}%
  \BibitemOpen
  \bibfield  {author} {\bibinfo {author} {\bibfnamefont {M.}~\bibnamefont
  {Barbieri}}, \bibinfo {author} {\bibfnamefont {M.}~\bibnamefont {Chotalia}},
  \bibinfo {author} {\bibfnamefont {J.}~\bibnamefont {Fraser}}, \bibinfo
  {author} {\bibfnamefont {L.-M.}\ \bibnamefont {Lavitas}}, \bibinfo {author}
  {\bibfnamefont {J.}~\bibnamefont {Dostie}}, \bibinfo {author} {\bibfnamefont
  {A.}~\bibnamefont {Pombo}},\ and\ \bibinfo {author} {\bibfnamefont
  {M.}~\bibnamefont {Nicodemi}},\ }\href@noop {} {\bibfield  {journal}
  {\bibinfo  {journal} {Biochem Soc Trans}\ }\textbf {\bibinfo {volume} {41}},\
  \bibinfo {pages} {508} (\bibinfo {year} {2013})}\BibitemShut {NoStop}%
\bibitem [{\citenamefont {Cortini}\ \emph {et~al.}(2016)\citenamefont
  {Cortini}, \citenamefont {Barbi}, \citenamefont {Caré}, \citenamefont
  {Lavelle}, \citenamefont {Lesne}, \citenamefont {Mozziconacci},\ and\
  \citenamefont {Victor}}]{Cortini2016}%
  \BibitemOpen
  \bibfield  {author} {\bibinfo {author} {\bibfnamefont {R.}~\bibnamefont
  {Cortini}}, \bibinfo {author} {\bibfnamefont {M.}~\bibnamefont {Barbi}},
  \bibinfo {author} {\bibfnamefont {B.~R.}\ \bibnamefont {Caré}}, \bibinfo
  {author} {\bibfnamefont {C.}~\bibnamefont {Lavelle}}, \bibinfo {author}
  {\bibfnamefont {A.}~\bibnamefont {Lesne}}, \bibinfo {author} {\bibfnamefont
  {J.}~\bibnamefont {Mozziconacci}},\ and\ \bibinfo {author} {\bibfnamefont
  {J.-M.}\ \bibnamefont {Victor}},\ }\bibfield  {journal} {\bibinfo  {journal}
  {Reviews of Modern Physics}\ }\textbf {\bibinfo {volume} {88}},\ \href
  {https://doi.org/10.1103/revmodphys.88.025002} {10.1103/revmodphys.88.025002}
  (\bibinfo {year} {2016})\BibitemShut {NoStop}%
\bibitem [{\citenamefont {Maeshima}\ \emph {et~al.}(2021)\citenamefont
  {Maeshima}, \citenamefont {Iida},\ and\ \citenamefont
  {Tamura}}]{Maeshima2021}%
  \BibitemOpen
  \bibfield  {author} {\bibinfo {author} {\bibfnamefont {K.}~\bibnamefont
  {Maeshima}}, \bibinfo {author} {\bibfnamefont {S.}~\bibnamefont {Iida}},\
  and\ \bibinfo {author} {\bibfnamefont {S.}~\bibnamefont {Tamura}},\ }\href
  {https://doi.org/10.1101/cshperspect.a040675} {\bibfield  {journal} {\bibinfo
   {journal} {Cold Spring Harbor Perspectives in Biology}\ }\textbf {\bibinfo
  {volume} {13}},\ \bibinfo {pages} {a040675} (\bibinfo {year}
  {2021})}\BibitemShut {NoStop}%
\bibitem [{\citenamefont {Brackley}\ \emph {et~al.}(2021)\citenamefont
  {Brackley}, \citenamefont {Gilbert}, \citenamefont {Michieletto},
  \citenamefont {Papantonis}, \citenamefont {Pereira}, \citenamefont {Cook},\
  and\ \citenamefont {Marenduzzo}}]{Brackley2021}%
  \BibitemOpen
  \bibfield  {author} {\bibinfo {author} {\bibfnamefont {C.~A.}\ \bibnamefont
  {Brackley}}, \bibinfo {author} {\bibfnamefont {N.}~\bibnamefont {Gilbert}},
  \bibinfo {author} {\bibfnamefont {D.}~\bibnamefont {Michieletto}}, \bibinfo
  {author} {\bibfnamefont {A.}~\bibnamefont {Papantonis}}, \bibinfo {author}
  {\bibfnamefont {M.~C.~F.}\ \bibnamefont {Pereira}}, \bibinfo {author}
  {\bibfnamefont {P.~R.}\ \bibnamefont {Cook}},\ and\ \bibinfo {author}
  {\bibfnamefont {D.}~\bibnamefont {Marenduzzo}},\ }\bibfield  {journal}
  {\bibinfo  {journal} {Nature Communications}\ }\textbf {\bibinfo {volume}
  {12}},\ \href {https://doi.org/10.1038/s41467-021-25875-y}
  {10.1038/s41467-021-25875-y} (\bibinfo {year} {2021})\BibitemShut {NoStop}%
\bibitem [{\citenamefont {Semeraro}\ \emph {et~al.}(2023)\citenamefont
  {Semeraro}, \citenamefont {Negro}, \citenamefont {Suma}, \citenamefont
  {Gonnella},\ and\ \citenamefont {Marenduzzo}}]{Semeraro2023}%
  \BibitemOpen
  \bibfield  {author} {\bibinfo {author} {\bibfnamefont {M.}~\bibnamefont
  {Semeraro}}, \bibinfo {author} {\bibfnamefont {G.}~\bibnamefont {Negro}},
  \bibinfo {author} {\bibfnamefont {A.}~\bibnamefont {Suma}}, \bibinfo {author}
  {\bibfnamefont {G.}~\bibnamefont {Gonnella}},\ and\ \bibinfo {author}
  {\bibfnamefont {D.}~\bibnamefont {Marenduzzo}},\ }\href
  {https://doi.org/10.1016/j.physa.2023.129013} {\bibfield  {journal} {\bibinfo
   {journal} {Physica A: Statistical Mechanics and its Applications}\ }\textbf
  {\bibinfo {volume} {625}},\ \bibinfo {pages} {129013} (\bibinfo {year}
  {2023})}\BibitemShut {NoStop}%
\bibitem [{\citenamefont {Kremer}\ and\ \citenamefont
  {Grest}(1990)}]{kremerDynamicsEntangledLinear1990}%
  \BibitemOpen
  \bibfield  {author} {\bibinfo {author} {\bibfnamefont {K.}~\bibnamefont
  {Kremer}}\ and\ \bibinfo {author} {\bibfnamefont {G.~S.}\ \bibnamefont
  {Grest}},\ }\href@noop {} {\bibfield  {journal} {\bibinfo  {journal} {J.
  Chem. Phys.}\ }\textbf {\bibinfo {volume} {92}},\ \bibinfo {pages} {5057}
  (\bibinfo {year} {1990})}\BibitemShut {NoStop}%
\bibitem [{\citenamefont {Weeks}\ \emph {et~al.}(1971)\citenamefont {Weeks},
  \citenamefont {Chandler},\ and\ \citenamefont {Andersen}}]{WCA}%
  \BibitemOpen
  \bibfield  {author} {\bibinfo {author} {\bibfnamefont {J.~D.}\ \bibnamefont
  {Weeks}}, \bibinfo {author} {\bibfnamefont {D.}~\bibnamefont {Chandler}},\
  and\ \bibinfo {author} {\bibfnamefont {H.~C.}\ \bibnamefont {Andersen}},\
  }\href {https://doi.org/10.1063/1.1674820} {\bibfield  {journal} {\bibinfo
  {journal} {The Journal of Chemical Physics}\ }\textbf {\bibinfo {volume}
  {54}},\ \bibinfo {pages} {5237} (\bibinfo {year} {1971})},\ \Eprint
  {https://arxiv.org/abs/https://pubs.aip.org/aip/jcp/article-pdf/54/12/5237/18874636/5237\_1\_online.pdf}
  {https://pubs.aip.org/aip/jcp/article-pdf/54/12/5237/18874636/5237\_1\_online.pdf}
  \BibitemShut {NoStop}%
\bibitem [{\citenamefont {Hinrichsen}(2000)}]{Hinrichsen2000}%
  \BibitemOpen
  \bibfield  {author} {\bibinfo {author} {\bibfnamefont {H.}~\bibnamefont
  {Hinrichsen}},\ }\href {https://doi.org/10.1080/00018730050198152} {\bibfield
   {journal} {\bibinfo  {journal} {Advances in Physics}\ }\textbf {\bibinfo
  {volume} {49}},\ \bibinfo {pages} {815–958} (\bibinfo {year}
  {2000})}\BibitemShut {NoStop}%
\bibitem [{\citenamefont {Henkel}\ \emph {et~al.}(2008)\citenamefont {Henkel},
  \citenamefont {Hinrichsen},\ and\ \citenamefont
  {L{\"u}beck}}]{HenkelHinrichsenLubeck2008}%
  \BibitemOpen
  \bibfield  {author} {\bibinfo {author} {\bibfnamefont {M.~M.}\ \bibnamefont
  {Henkel}}, \bibinfo {author} {\bibfnamefont {H.}~\bibnamefont {Hinrichsen}},\
  and\ \bibinfo {author} {\bibfnamefont {S.}~\bibnamefont {L{\"u}beck}},\
  }\href@noop {} {}\ (\bibinfo {year} {2008})\BibitemShut {NoStop}%
\bibitem [{\citenamefont {Mao}\ \emph {et~al.}(2011)\citenamefont {Mao},
  \citenamefont {Zhang},\ and\ \citenamefont {Spector}}]{MAO2011295}%
  \BibitemOpen
  \bibfield  {author} {\bibinfo {author} {\bibfnamefont {Y.~S.}\ \bibnamefont
  {Mao}}, \bibinfo {author} {\bibfnamefont {B.}~\bibnamefont {Zhang}},\ and\
  \bibinfo {author} {\bibfnamefont {D.~L.}\ \bibnamefont {Spector}},\ }\href
  {https://doi.org/https://doi.org/10.1016/j.tig.2011.05.006} {\bibfield
  {journal} {\bibinfo  {journal} {Trends in Genetics}\ }\textbf {\bibinfo
  {volume} {27}},\ \bibinfo {pages} {295} (\bibinfo {year} {2011})}\BibitemShut
  {NoStop}%
\bibitem [{\citenamefont {Erdel}(2023)}]{Erdel2023}%
  \BibitemOpen
  \bibfield  {author} {\bibinfo {author} {\bibfnamefont {F.}~\bibnamefont
  {Erdel}},\ }\href {https://doi.org/10.1016/j.sbi.2023.102597} {\bibfield
  {journal} {\bibinfo  {journal} {Current Opinion in Structural Biology}\
  }\textbf {\bibinfo {volume} {80}},\ \bibinfo {pages} {102597} (\bibinfo
  {year} {2023})}\BibitemShut {NoStop}%
\end{thebibliography}%







\end{document}